\documentclass[aps,preprint,groupedaddress,showpacs]{revtex4}
\usepackage{graphicx}

\newcommand{\et}{\epsilon_{14}}
\newcommand{\apjl}{Astrophys.\ J.}

\begin{document}
\bibliographystyle{apsrev}

\preprint{}

\title{High-Energy Neutrinos from Gamma Ray Bursts}

\author{Charles D. Dermer$^1$}\author{Armen Atoyan$^2$}
\affiliation{$^1$Code 7653, Naval Research Laboratory, Washington, 
DC 20375-5352 \\
$^2$CRM, Universite de Montreal, Montreal H3C 3J7, Canada}

\date{\today}

\begin{abstract}
We treat high-energy neutrino production in GRBs. Detailed
calculations of photomeson neutrino production are presented for the
collapsar model, where internal nonthermal synchrotron radiation is
the primary target photon field, and the supranova model, where
external pulsar-wind synchrotron radiation provides important
additional target photons. Detection of $\gtrsim 10\,\rm TeV$
neutrinos from GRBs with Doppler factors $\gtrsim 200$, inferred from
$\gamma$-ray observations, would support the supranova model.
Detection of $\lesssim 10 \,\rm TeV$ neutrinos is possible for
neutrinos formed from nuclear production.  Only the most powerful
bursts at fluence levels $ \gtrsim 3\times 10^{-4}\,\rm erg \,
cm^{-2}$ offer a realistic prospect for detection of $\nu_\mu$.

\end{abstract}

\maketitle

Two leading scenarios for the nature of the sources that power
long-duration GRBs are the collapsar \cite{woo} and supranova (SA)
\cite{vs} models. The core of a massive star collapses directly
to a black hole in the collapsar model, but only after an
episode of neutron-star activity in the SA model \cite{kg02}. This
delay means that a shell of enriched material, which could help
explain observations of X-ray features, surrounds the GRB source in
the SA model \cite{vs}.  The colliding shells thought to operate in
the collapsar model take place at distances $\Gamma^2 c \Delta t \sim
3\times 10^{15}\Gamma^2_{300}\Delta t($s) cm from the central ejection
source, where $\Gamma = 300\Gamma_{300}$ is a typical wind Lorentz
factor and $\Delta t$ is the time between shell ejection events
\cite{dm98}. GRB pulses of $\sim 0.1$-10 s durations and separations
are typical \cite{nor}, but even for $\Delta t \sim 1$ ms, the shell
collisions take place far outside the photospheric radii of likely GRB
stellar progenitors. The most important radiation field for photomeson
neutrino production in the collapsar scenario is thus the internal
synchrotron radiation \cite{wb}.

The presence of a pulsar wind (PW) within an expanding supernova
remnant (SNR) shell in the SA model means that the radiation
environments in the collapsar and SA models are vastly different
\cite{kg02}.  To maintain stability against prompt collapse to a black 
hole, the neutron star must rotate with periods near 1 ms \cite{vs}.
The neutron star radiates electromagnetic, leptonic, and possibly also
hadronic energy in the form of a powerful relativistic `cold'
magnetized wind consisting of quasi-monoenergetic $e^{+}$-$e^{-}$
pairs and ions.  The ordered flow of the wind is disrupted at the wind
shock formed within the SNR shell. Here we consider only the emission
from quasi-monoenergetic leptons that are injected at the PW shock
which provide, primarily through synchrotron losses, the main external
photon target for photomeson interactions.  
The quasi-thermal radiation field of the SNR
shell is neglected, as well as the nonthermal synchrotron and
Compton-scattered radiations from particles accelerated at the PW
shock.

We find that the external lepton-wind synchrotron radiation alone
improves prospects for neutrino detection by orders of magnitude over
values calculated in a standard external-shock/GRB blastwave scenario
\cite{der02a} that explains well the phenomenology of GRBs with smooth
fast-rise/slow decay $\gamma$-ray light curves. The presence of the
external field can increase the number of detectable neutrinos by an
order of magnitude or more over a colliding shell scenario that is
generally invoked to explain highly variable GRB light curves.

Here we report detailed calculations of photomeson neutrino
production for the collapsar and SA models that take into account
nonthermal proton injection followed by photomeson energy loss, which
is computed numerically by adapting our photo-hadronic model for
blazar jets \cite{ad01}.  We also calculate the corresponding cutoff
spectral energy $\epsilon_{\gamma\gamma}$ due to $\gamma\gamma$
pair-production attenuation.  High-energy neutrino observations of
very bright GRBs, especially if combined with data from the {\it
Gamma-ray Large Area Space Telescope} ({\it GLAST}) or other gamma-ray
detectors, can test the collapsar and SA models, as we now show.

The detection efficiency in water or ice of ultrarelativistic
upward-going muon neutrinos ($\nu_\mu$) with energies
$\epsilon_\nu = 10^{14}\et$ eV is $P_{\nu\mu} \cong
10^{-4}\et^\chi$,  where $\chi = 1$ for $\et <1$,
and $\chi =0.5$ for $\et>1$ \cite{ghs95}. 
For a neutrino fluence spectrum parameterized by $\nu \Phi_\nu =
10^{-4}\phi_{-4}\et^{\alpha_\nu}$ erg cm$^{-2}$, the number of
$\nu_\mu$ detected with a km-scale $\nu$ detector 
such as IceCube with area $A_\nu = 10^{10}A_{10}$ cm$^2$ is
therefore
\begin{equation}
N_\nu(\geq \et) \approx \int_{\epsilon_\nu}^{\infty} {\rm d} \epsilon_1
  \;{\nu\Phi_\nu \over
\epsilon_{1}^{2}}\;P_{\nu\mu}A_\nu \simeq
0.6\;{\phi_{-4} A_{10}\over
{{1\over 2} - \alpha_\nu}}\cases{1+({1\over 2\alpha_\nu} - 1)
(1-\et^{\alpha_\nu}) \; ,&
for $\et<1$ \cr\cr \et^{{\alpha_\nu} -1/2}\; , & for
$\et > 1$ .\cr}\;\;
\label{Nnu}
\end{equation}
For a $\nu\Phi_\nu$ spectrum with $\alpha_\nu \simeq 0$, the number of
$\nu_\mu$ to be expected are $N_\nu \simeq 1.2\phi_{-4} A_{10}
(1+{1\over 2} \ln \et^{-1})$ for $\et<1$, and $N_\nu \simeq
1.2\phi_{-4} A_{10} /\sqrt{\et}$ for $\et >1 $.  If the nonthermal
proton energy injected in the proper frame is comparable to the
radiated energy required to form GRBs with hard X-ray/soft
$\gamma$-ray fluences $\gtrsim 10^{-4}$ ergs cm$^{-2}$, then extremely
bright GRBs are required to leave any prospect for detecting $\nu_\mu$
with km-scale neutrino detectors. About 2-5 GRBs per year are expected
with hard X-ray/soft $\gamma$-ray fluence $> 3\times 10^{-4}$ ergs
cm$^{-2}$ \cite{bri99}.

Because GRB blast waves are believed to accelerate nonthermal
electrons, it is probable that they also accelerate nonthermal
hadrons. Coincidence between the GRB power radiated at hard X-ray/soft
$\gamma$-ray energies within the GZK radius and the power required to
account for super-GZK ($\gtrsim 10^{20}$ eV) cosmic rays suggests that
GRBs might be the progenitor sources of ultra-high energy ($\gtrsim
10^{19}$ eV) cosmic rays, cosmic rays above the ankle of the cosmic
ray spectrum, and GeV-TeV cosmic rays \cite{vw}\cite{der02a}.

In our calculations, we inject protons with a number spectrum $\propto
\epsilon_p^{-2}$ at comoving proton energies $\epsilon_p >
300\Gamma_{300}$ GeV up to a maximum proton energy determined by the
condition that the particle Larmor radius is smaller than both the
size scale of the emitting region and the photomeson energy-loss
length \cite{ad01}.  The observed synchrotron spectral flux in the
prompt phase of the burst is parameterized by the expression $F(\nu) \propto
\nu^{-1} (\nu/\nu_{br})^{\alpha}$, where $h\nu_{br}=300$ keV,
$\alpha = -0.5$ above $\nu_{br}$, and $\alpha = 0.5$ when $30\, {\rm
keV} \leq h\nu \leq h \nu_{br}$. At lower energies, $\alpha= 4/3$.
The observed total hard X-ray/soft $\gamma$-ray photon fluence $
\Phi_{tot} \cong t_{dur}\int_0^\infty d\nu F(\nu )$, where $t_{dur}$
is the characteristic duration of the GRB. In our calculations we
assume a source at redshift $z=1$, and let $\Phi_{tot} =3\times
10^{-5} \,\rm erg \; cm^{-2}$. Two or three GRBs should occur each month
above this fluence level.

We inject a total amount of accelerated proton energy $E^\prime = 4\pi
d_L^2\Phi_{tot} \delta^{-3} (1+z)^{-1}$ into the comoving frame of the GRB
blast wave. Here $\delta$ is the Doppler factor and $d_L$ is the
luminosity distance. The energy deposited into each of $N_{sp}$
light-curve pulses (or spikes) is therefore $E_{sp}^\prime =
E^\prime/N_{sp}\,$ ergs.  We assume that all the energy
$E_{sp}^\prime$ is injected in the first half of the time interval of
the pulse, which effectively corresponds to a characteristic
variability time scale $t_{var} = t_{dur}/2N_{sp}$.  The proper width
of the radiating region forming the pulse is $\Delta R^\prime \cong t_{var}
c\delta/(1+z)$, from which the energy density of the synchrotron
radiation can be determined \cite{ad01}.  We set the GRB prompt
duration $t_{dur} = 100\,$s, and let $N_{sp} = 50$, corresponding to
$t_{var} = 1\,\rm s$.  The magnetic field is determined by assuming
equipartition between the energy densities of the magnetic field and
the electron energy.

Fig.\ 1 shows the optical depth $\tau_{\gamma\gamma}$ for
$\gamma\gamma$ pair production attenuation as a function of observer
photon energy, calculated for $\delta = 100, 200,$ and 300.  Curves 1,
2, and 3 show the $\gamma\gamma$ opacity from internal synchrotron
radiation only, corresponding to the collapsar scenario. The rapid
decrease of $\tau_{\gamma\gamma}$ with increasing $\delta$ is
explained by the rapid decrease of the energy density of the internal
radiation in the comoving frame as $u_{syn}^\prime \simeq
L_{syn}^\prime /2\pi R^{\prime 2} c \propto (\nu
F_\nu)_{obs}/\delta^6$, which implies a rapid decline (approximately,
$\tau_{\gamma\gamma} \propto \delta^{-4}$, depending precisely on the
radiation spectrum) of the $\tau_{\gamma\gamma}$ opacity in the
synchrotron field.  Relativistic flows with $\delta
\gtrsim 100$ are required to explain observations of $> 100$ MeV
$\gamma$ rays with EGRET \cite{bar}. {\it GLAST} observations may
imply more stringent limits on $\delta$.

An external radiation field given by the expression $\nu L_\nu
\propto\nu^{1/2} \exp(-\nu/\nu_{ext})$, with $ h\nu_{ext} = 0.1$ keV,
is assumed to be present in the SA model \cite{kg02}.  The intensity
of this field is determined by the assumption that the integral power
$L_{ext} =\int_0^\infty L_\nu \rm d \nu$ is equal to the power 
of the pulsar wind $L_{pw}\approx (10^{53}\,{\rm erg})/t_{delay}$,
assuming that a total of $\approx 10^{53} \,\rm erg$ of pulsar
rotation energy is radiated during the time $t_{delay}$ (which is here
set equal to 0.1 yr) from the rotating supramassive neutron star
before it collapses to a black hole.  The energy $h\nu_{ext}\simeq 0.1
\,\rm keV$ is the characteristic energy of synchrotron radiation
emitted by electrons (of the pulsar wind) with Lorentz factors
$\gamma_{pw}\sim 3\times 10^4$ in a randomly ordered magnetic field of
strength $\approx 10$ G. The radius $R$ is determined by assuming
$v=0.05c$ is the mean speed of the SNR shell, and the external photon
energy density $\propto L/2\pi R^2$.  The thin solid curve is the
combined opacity in a SA-model pulse when $\delta = 100$. In this
case, the pulses will be highly attenuated above $\approx 300$ MeV.

Fig.\ 2 shows the total $\nu_\mu$ fluences expected from a model GRB
with $N_{sp}=50$ pulses. 
The thin curves show collapsar model results at $\delta = 100,$ 200,
and 300. The expected numbers of $\nu_\mu$ that a km-scale detector
such as IceCube would detect are $N_\nu = 3.2\times 10^{-3}$,
$1.5\times 10^{-4}$, and $1.9\times 10^{-5}$, respectively.  (The
effect of neutrino flavor oscillations could reduce these
numbers by a factor $\approx 2$.)  There is no prospect to detect
$\nu_\mu$ from GRBs at these levels.  The heavy solid and dashed
curves in Fig.\ 2 give the SA model predictions of $N_\nu = 0.009$ for
both $\delta = 100$ and $\delta = 300$. The equipartition magnetic
fields are 1.9 kG and 0.25 kG, respectively. The external radiation
field in the SA model makes the neutrino detection rate insensitive to
the value of $\delta$ (as well as to $t_{var} \gtrsim 0.1\,\rm s$, as
verified by calculations), but there is still little hope that a
km$^3$ detector could detect such GRBs.

Neutrino production efficiency would improve in the collapsar and SA
models if $t_{var} \sim 1\,$ms and $N_{sp} = 5\times 10^4$ to provide
the same total fluence. We obtain $N_\nu \cong 0.027$, 0.012, and
0.0046 for $\delta = 100$, 200, and 300, respectively. Such narrow
spikes are, however, then nearly opaque to gamma rays, with
$\tau_{\gamma\gamma} = 1$ at $E_\gamma \cong 80$ MeV when $\delta =
200$, and at $E_\gamma \cong 500$ MeV when $\delta = 300$. 
When $t_{var}\gg 1 \,\rm s$, the GRB blast wave is optically
thin to its internal synchrotron radiation at multi-GeV energies. In
the absence of an external radiation field, the produced fluxes
of neutrinos are not however detectable.

Neutrino detection from GRBs can be assured only if
the number of background counts 
\begin{equation}
B(\geq \et) \cong \int d\Omega \int dt \int_{\epsilon_\nu}^\infty
 d\epsilon_1 \; 
{F_{\nu}^{atm} \over \epsilon_{1}^{2}}\;P_{\nu\mu} A_\nu \simeq 
5 \times 10^{-8} t_2 A_{10} \left({\theta\over 1^\circ}\right)^2 
\int_{\et}^\infty dx \, x^{-\beta+\chi-2} \ll 1.
\label{B}
\end{equation}
Here the chosen time window has duration $t_w = 10^2t_2$ s, $\theta$
is the half-opening angle, and the cosmic-ray induced atmospheric
neutrino background flux at the nadir is $F_{\nu}^{atm} \approx
8\times 10^{-11} \et^{-\beta}$ ergs cm$^{-2}$ s$^{-1}$ sr$^{-1}$, with
$\beta \cong 1.7$ for $\et < 1$, and $\beta = 2$ for $\et > 1$
\cite{ghs95}.  This expression places a lower bound on the energy
above which background counts may be neglected, namely that the
neutrino energy $\epsilon_\nu \gg \epsilon_{min} \cong 3.6 (t_{2}
A_{10})^{0.59} (\theta/1^\circ)^{1.18}$ GeV in order to have $B
\ll 1 $.  Note that IceCube's angular resolution is predicted
to be $\sim 1^\circ$ at TeV energies, $0.5^\circ$ at 50-100 TeV
energies, and $0.6^\circ$ at PeV energies \cite{ghs95}.  Therefore a
detection of only 2 neutrinos during the time interval of the prompt
phase of a GRB, especially if at energies $\epsilon_\nu \gtrsim 10
\,\rm TeV$ predicted by photohadronic interactions, will be highly
significant.
 
The importance of low-energy (10 GeV - 10 TeV) sensitivity in neutrino
experiments is connected with the possibility to probe jet models with
nuclear interactions.  One possible scenario is a beam-on-target model
where the GRB protons pass through and occasionally collide with
particles in a dense target, namely the SNR shell in the SA scenario
\cite{kat}. Neutrinos created from the decay of mesons formed in
interactions with SNR shell particles with mean atomic weight $A$ are
beamed into an angle no smaller than the opening angle of the GRB
outflow. Thus the neutrino fluence is at most comparable to the hard
X-ray/soft $\gamma$-ray fluence of a GRB multiplied by the conversion
efficiency $\eta_p= t_{sh}/3t_{pA}$ of protons to neutrinos, again
assuming as before that the GRB hadronic energy is determined by
$\Phi_{tot}$.  The factor of 3 accounts for the fraction of secondary
energy emitted in the form of $\nu_\mu$.  Here the available time for
the interaction of relativistic protons passing through the SNR shell
is $t_{sh} \cong 3.3 \times 10^{4} f_{-1} R_{16}
\,\rm s$, assuming a uniform shell with a characteristic thickness
$0.1 f_{-1} R$ and radius $R=10^{16}R_{16}$.  The $pA
\rightarrow\pi^{\pm,0}$ energy-loss time scale $t_{pA}$ in the
stationary frame is given by $t^{-1}_{pA}
\cong K_p n_A\sigma_{pA} c\simeq 4.3 \times 10^{-7}
 m_{SNR}/( A^{1/3} R_{16}^3 f_{-1}) \,\rm s^{-1}$, given an
inelasticity $K_p\simeq 0.5$ and a strong interaction cross section
$\sigma_{pA} \cong 30 A^{2/3}$ mb.  Thus
$\eta_{p} \simeq 5 \times 10^{-3} m_{SNR} R_{16}^{-2} A^{-1/3}$. These
neutrinos are formed at energies $\simeq 0.05
\Gamma m_p c^2 \simeq 15\Gamma_{300}$ GeV. 
Neutrinos formed through beam/target interactions are therefore
relatively low energy, and are not expected from GRBs unless $R_{16}
\ll 1$, in which case the SNR shell would be very Thomson thick.

Another nuclear interaction scenario
is where relativistic particles in the GRB blast wave collide with
other particles in the ejecta. The relativistic particles are formed
either by sweeping up and isotropizing particles captured from the
external medium \cite{ps}, or by accelerating particles at external or
internal shocks. The $\nu_\mu$-production efficiency $\eta_{pp}\cong
t^\prime_{ava}/3t^\prime_{pp}$, where $t^\prime_{ava}
\lesssim R/\Gamma c$ is the available time in the comoving frame, 
 $t^{\prime ~-1}_{pp} = n_{p}^{\prime}\sigma_{pp} c = E_{iso}
\sigma_{pp}c/(4\pi R^2 \Delta R^\prime \Gamma m_p c^2)$ is the inverse
of the proper-frame nuclear-collision time scale, and $E_{iso} =
10^{52}E_{52}$ ergs is the apparent isotropic GRB energy
release. Because $\Delta R^\prime \sim \Gamma c t_{var}$, we find
$\eta_{pp} \lesssim 6\times 10^{-4} E_{52}/ [R_{16}\Gamma_{300}^{2}
t_{var}({\rm s})]$. Secondary neutrinos are formed at energies
$\approx 0.05\Gamma^2m_pc^2 \cong 4\Gamma^2_{300}$ TeV. Only GRBs with
large values of $E_{52}$ or small values of $\Gamma_{300}$, $R_{16}$, and
$t_{var}$(s) provide reasonable efficiencies for $\nu_\mu$ production,
and these would exhibit strong $\gamma\gamma$ attenuation
with comparable fluences in $\nu_\mu$ and 
reprocessed electromagnetic radiation.

Thus detection of a few neutrinos by a km-scale detector generally
requires an extremely bright event at the level reaching $\sim
10^{-3}\,\rm erg\; cm^{-2}$.  Note in this regard that a model
recently proposed \cite{gg02} for neutrino emission from the $pp$
interaction of PW protons with the SNR shell during the time
$t_{delay}$ between supernova and GRB events neglects the expected
flux of accompanying electromagnetic radiation, such that a pre-GRB
source in the SA scenario would be an extremely bright source
significantly {\it before} the GRB event, which would be hard to miss
in all-sky X-ray and $\gamma$-ray observations.  A recent study of
neutrino production during the prompt GRB phase in the SA model
\cite{gg02a} finds similar estimates for the number of expected
neutrinos which, however, are predicted to be mostly at much higher
enegies, $> 1000 \,\rm TeV$. Moreover, photomeson interactions of
protons on the internal synchrotron radiation field as well as
$\gamma\gamma$ attenuation are not studied in Ref.\
\cite{gg02a,rmw02}, which is important to discriminate between
different GRB models.

Another way to improve high-energy neutrino and UHECR
model production efficiency is to assume that the radiating particles
in the GRB blast waves are dominated by hadrons. As Fig.~2 shows, the
integrated level of $\nu_{\mu}$ fluence and therefore of secondary
gamma-rays for the assumed injection of relativistic protons is at
the level $\lesssim 3\times 10^{-6}$ erg cm$^{-2} \sim 0.1
\Phi_{tot}$. It would therefore still be possible to assume acceleration of
protons with a power up to a factor $\approx 10$ higher than in
Fig.~2, which would increase the number of $\nu_\mu$ expected for
IceCube to $\sim 0.1$ in the SA model. The detection of a few
neutrinos would then be possible in the prompt phase of GRBs at the flux
level $\Phi_{tot}\gtrsim 3\times 10^{-4} \,\rm erg/cm^2$. It is
important that from photomeson interactions only neutrinos above 10-20
TeV are to be expected, while a detection of lower energy neutrinos
would significantly favor models with nuclear interactions in either
the SA or the collapsar scenarios.

In this letter, we have shown that the collapsar model is less
efficient for neutrino production than the supranova model, as it
lacks the external PW synchrotron radiation field. Neutrino
predictions with the collapsar model are very sensitive to $\delta$,
and are most favorable when the $\sim 100$ MeV - GeV $\gamma\gamma$
opacity is large.  An inverse relation between the number of
detectable neutrinos and cutoff spectral energy is found for a
collapsar model. Neutrino production efficiency is relatively
insensitive to $\delta$ in the SA model, though it does depend on model
parameters of the PW synchrotron emission.  There is no prospect for
neutrino detection in the collapsar model when $\delta \gtrsim 200$.
More optimistic estimates from the viewpoint of detecting GRB
neutrinos could be found in proton-dominated GRB models.  Without this
hypothesis only the brightest GRBs can be expected to be detected with
both high-energy $\gamma$-ray and neutrino detectors. Detection of
high-energy neutrinos from GRBs would therefore have far-reaching
impact on the collapsar and SA scenarios for GRBs, the hypothesis that
GRBs sources can be powerful accelerators of ultrarelativistic CRs,
and for our understanding of the origin of GRB radiation.

\vskip0.2in
\noindent AA appreciates the support and hospitality 
of the NRL High Energy Space Environment Branch during his visit when
 this work was initiated. The work of CD is supported by the Office
 of Naval Research and NASA GLAST science investigation grant DPR \#
 S-15634-Y.

\bibliography{your bib file}

\begin{figure}
{\includegraphics{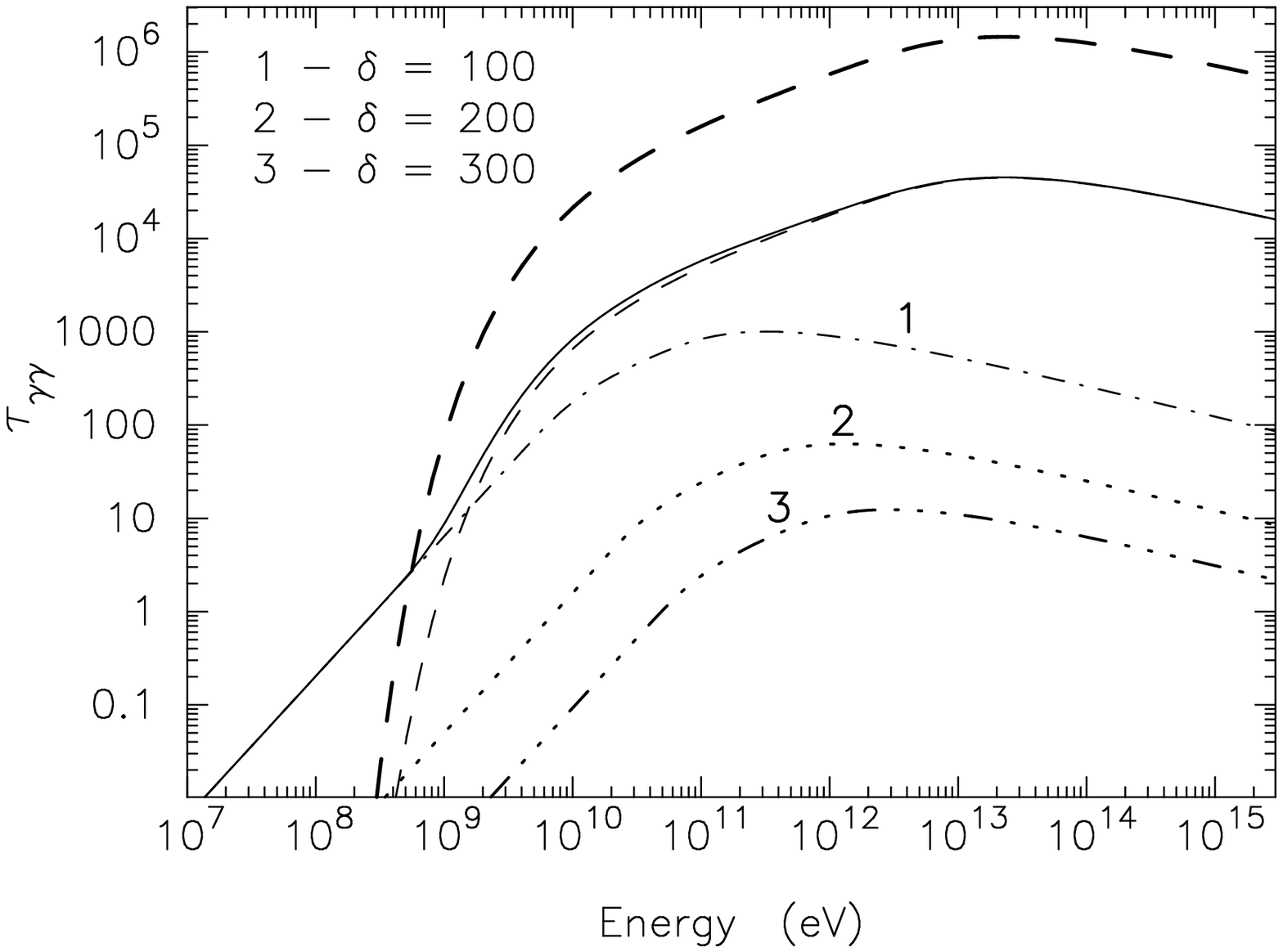}}
\caption[]{
Opacity $\tau_{\gamma\gamma}$ to pair-production attenuation for
$\gamma$ rays which are detected with observer energy given on the
abscissa.  The radiating region moves with Doppler factor $\delta$
with respect to the observer. The internal synchrotron radiation field
is typical of GRBs observed with BATSE, and the injected energy
corresponds to a measured fluence of $3\times 10^{-5}$ ergs cm$^{-2}$,
which is radiated in 50 equal pulses.  The heavy and light dashed
curves give the $\gamma\gamma$ opacity through the external radiation
field over the size scale of the supernova remnant and the pulse
emitting region, respectively. The thin solid curve gives the total
opacity for a model pulse in the external shock/SA model.   }
\label{Fig1}
\end{figure}

\begin{figure}
{\includegraphics{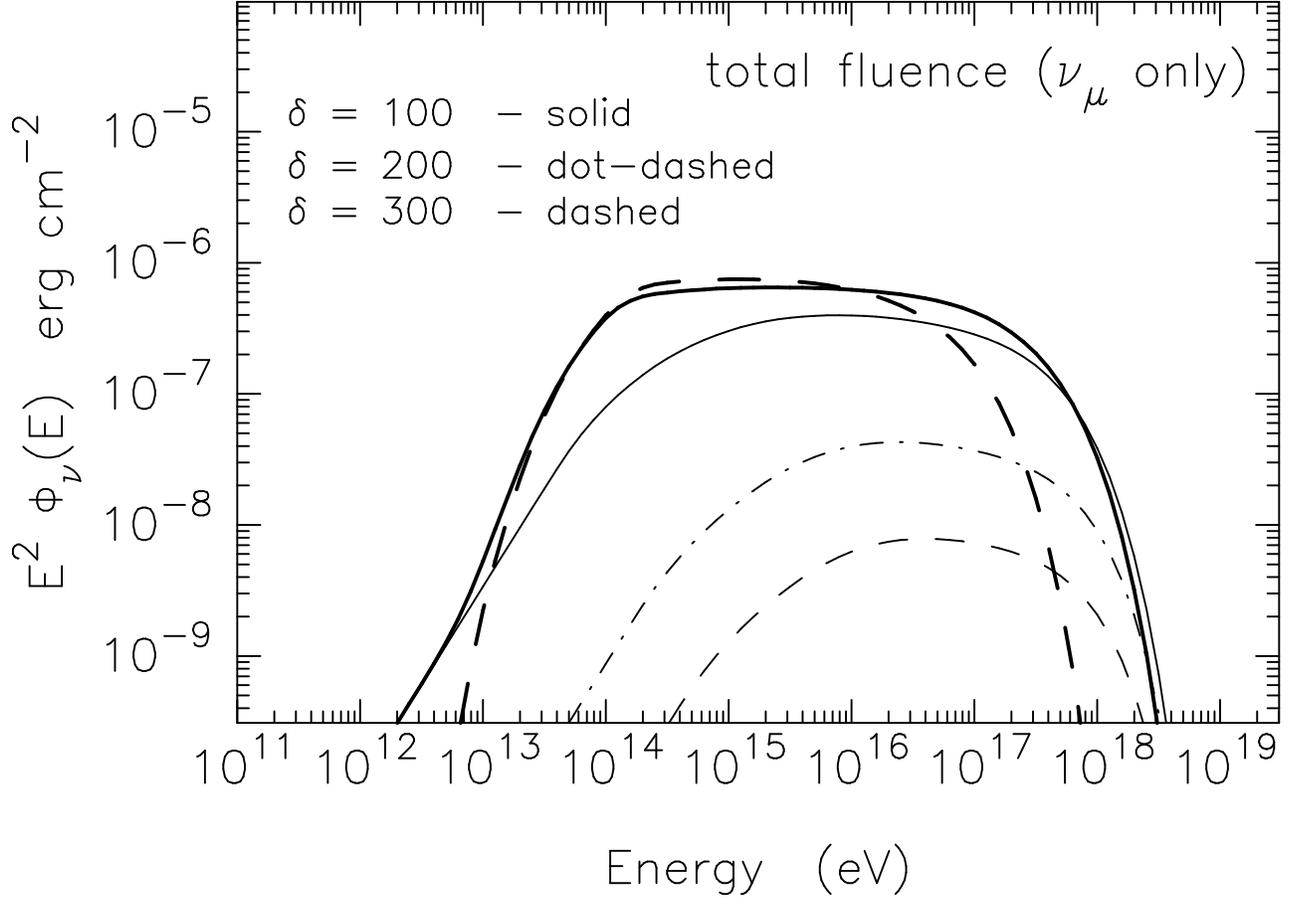}}
\caption{
Energy fluence of photomeson muon neutrinos for a model GRB. The thin
curves show collapsar model results where only the internal
synchrotron radiation field provides a source of target photons.  The
thick curves show the $\delta = 100$ and 300 results for the SA-model
calculation, which includes the effects of an external pulsar wind
radiation field.  }
\label{Fig2}
\end{figure}

\end{document}